\documentclass[aps,prb,showpacs,twocolumn]{revtex4-1}
\usepackage{bm,color,amsmath,amssymb,mathrsfs,latexsym,graphicx,psfrag}
\newcommand{\bs}[1]{\boldsymbol{#1}}

\newcommand{\be}{\begin{equation}}
\newcommand{\ee}{\end{equation}}

\def\be{\begin{equation}}
\def\ee{\end{equation}}
\def\bea{\begin{eqnarray}}
\def\eea{\end{eqnarray}}

\def\C60{A$_x$C$_{60}$}

\def\HgCu3{HgCa$_2$Cu$_3$O$_{8+y}$}
\def\HgCu4{HgBa$_2$Ca$_3$Cu$_4$O$_{10+y}$}
\def\TlCu{Tl$_2$Ba$_2$CuO$_{6+\delta}$}
\def\TlCu3{Tl$_2$Ba$_2$Ca$_2$Cu$_3$O$_{10+y}$}
\def\TlCu4{Tl$_2$Ba$_2$Ca$_3$Cu$_4$O$_{12+y}$}

\def\BiCu3{Bi$_2$Sr$_2$Ca$_{2}$Cu$_3$O$_y$}

\def\8LSCO{La$_{1.88}$Sr$_{.12}$CuO$_4$}
\def\110LNSCO{La$_{1.5}$Nd$_{0.4}$Sr$_{0.1}$CuO$_{4}$}
\def\stage4LCO{La$_{2}$CuO$_{4+\delta}$}
\def\Y248{YBa$_2$Cu$_4$O$_8$}

\def\NbSe2{NbSe$_2$}
\def\TaSe2{TaSe$_2$}
\def\TiSe2{TiSe$_2$}

\begin{document}

\title{ Robustness of $s$-wave Pairing in Electron-Overdoped
$\text{A}_{1-y}\text{Fe}_{2-x}\text{Se}_2$}
\author{Chen Fang${}^1$}
\author{Yang-Le Wu${}^{2}$}
\author{Ronny Thomale${}^2$}
\author{B.~Andrei Bernevig${}^2$}
\author{Jiangping Hu${}^{1,3}$}
\affiliation{${}^{1}$Department of Physics, Purdue University, West
Lafayette, Indiana 47907, USA}
\affiliation{${}^{2}$Department of Physics,
Princeton University, Princeton,
  NJ 08544}
\affiliation{${}^{3}$Beijing National Laboratory for Condensed Matter
Physics, Institute of Physics, Chinese Academy of Sciences, Beijing
100080, China}

\begin{abstract}
  Using self consistent mean field and functional renormalization group
  approaches we show that $s$-wave pairing symmetry is robust in the heavily
  electron-doped iron chalcogenides $(\text{K, Cs}) \text{Fe}_{2-x}\text{Se}_2 $. This is because in
  these materials the leading
  antiferromagnetic (AFM) exchange coupling is between
  next-nearest-neighbor (NNN) sites while the nearest neighbor (NN)
  magnetic exchange coupling is  ferromagnetic (FM).  This is different from the iron pnictides, where the
  NN magnetic exchange coupling is AFM and leads to strong
  competition between $s$-wave and $d$-wave pairing in the electron
  overdoped region. Our finding of a robust $s$-wave pairing in $(\text{K, Cs}) \text{Fe}_{2-x}\text{Se}_2$ differs from the $d$-wave pairing result obtained by
  other theories where non-local bare interaction terms and the NNN
  $J_2$ term are underestimated.  Detecting the pairing
  symmetry in $(\text{K, Cs}) \text{Fe}_{2-x}\text{Se}_2 $ may hence provide
  important insights regarding the mechanism of superconducting
  pairing in  iron based superconductors.
\end{abstract}

\maketitle

\section{Introduction}
\label{sec:intro}
The recent discovery of a new family of iron-based superconductors
$\text{A(K,Cs,Rb)}_y\text{Fe}_{2-x}\text{Se}_2$\cite{Guo2010,Fang2010,Liu2011}
has initiated a new round of research in this field. Remarkably, this
new family shows distinctly different properties from other pnictide
families: the compounds are heavily electron doped, but their
superconducting transition temperatures are high, at more than $40$ K. For
comparison, such large $T_c$'s can only be reached in the optimally
doped 122 iron pnictides~\cite{basovchubu}. Importantly, both angle-resolved
photoemission spectroscopy
(ARPES)\cite{ZhangY2010,WangXP2011,Mou2011} and LDA
calculations\cite{Cao2010,ZhangL2009,Yan2010} show the presence of
only electron Fermi pockets located at the $M$ point
of the folded Brillouin zone (BZ). (Some signature of possible density of
states at the $\Gamma$ point is still under current debate; in any case this
pocket, if present, is assumed to be very flat and shallow). ARPES
experiments have also reported large isotropic superconducting gaps at
these pockets\cite{ZhangY2010,WangXP2011,Mou2011}. The absence of
hole pockets around the $\Gamma$ point of the BZ provides
a new arena of Fermi surface topology to investigate the pairing
symmetries and mechanisms of superconductivity proposed for iron-based
superconductors from a variety of different
approaches\cite{seo2008,Fang2008,
  Wang2008a,Mazin2008,chubukov2008,Maier2008d,Thomale2009,Thomale2011,
  si,Berciu2008,Chen2008f,dai2008j,Kou2008,Haule2008,Haule2009,Wu2009,
  Daghofer2008,Mishra2009,Lee2008a,Cvetkovic2009,Kuroki2008,Wang2009,grasermaierhirschscala2,grasermaierhirschscala,
elbi}.

So far, the majority of theories for the pairing symmetry of
iron-based superconductors are based on weak coupling
approaches\cite{
  Wang2008a,Mazin2008,chubukov2008,Maier2008d,Thomale2009,Thomale2011,
  Daghofer2008,Mishra2009,Lee2008a,Cvetkovic2009,Kuroki2008,Wang2009,kangtesa,maitichubu}. Although
there are discrepancies, the theories based on these approaches have
reached a broad consensus regarding the pairing symmetries
in iron-based superconductors: for optimally hole doped
iron-pnictides, for example, $\text{Ba}_{0.6}\text{K}_{0.4}\text{Fe}_2\text{As}_2$, an extended
$s$-wave pairing symmetry, called $s^\pm$, is favored\cite{Mazin2008}
(the sign of the order parameter changes between hole and electron
pockets as potentially detectable through neutron scattering~\cite{maiergraserhirsch}), as a result of repulsive interband interactions and nesting
between the hole and electron pockets. For extremely hole-doped
materials, such as $\text{KFe}_2\text{As}_2$, the absence of electron pockets can
lead to a $d$-wave pairing symmetry\cite{Thomale2011} with a low transition temperature; for electron
doped materials such as $\text{Ba}_2 \text{Fe}_{2-x}\text{Co}_x \text{As}_2$, the anisotropy
of the superconducting gap around the electron pockets in the $s^\pm$
state grows for larger electron doping and eventually the SC gap
develops nodes around the electron pockets due to the weakening of the
nesting condition and the increase of $d_{xy}$ orbital weight at the
electron pocket Fermi surfaces\cite{Thomale2010asp,stanevnikotesa}.  Finally, in the limit of the heavily
electron doped case when the hole pockets vanish and only the electron
pockets are left, the $d$-wave pairing symmetry may be favored
again\cite{Thomale2011,cth,maitikorsh}. The iron chalcogenide $\text{A}_y\text{Fe}_{2-x}\text{Se}_2$ belongs to the latter
category and many theories based on weak coupling approaches have
suggested that the pairing symmetry should be
$d$-wave as possibly detectable through characteristic impurity scattering\cite{Youy2011,Maier2011,balatsi}.

A complementary approach based on strong coupling likewise predicts an
$s$-wave pairing symmetry in the iron pnictides.  Two of us showed that
the pairing symmetry is determined mainly by the next-nearest-neighbor
(NNN) AFM exchange coupling $J_2$ together with a
renormalized narrow band width\cite{seo2008,Parish2008c}.
The superconducting gap is close to a $ \cos k_x \cos k_y$ form in
momentum space (higher harmonic contributions are neglected in this
approach). This result is model independent as long as the dominating
interaction is $J_2$ and the Fermi surfaces are located close to the
$\Gamma$ and M points in the folded BZ. The $\cos k_x
\cos k_y$ form factor changes sign between the electron and hole
pockets in the BZ. It resembles the order parameters of
$s^\pm$ proposed from weak-coupling arguments\cite{Mazin2008}.

%Most recently, we have also shown that the $s$-wave pairing
%symmetry is rather robust in the strongly coupling limit for a
%multi-orbital checkerboard model if a small antiferromagnetic $J_2$
%is added\cite{Lu2010}.

The $J_2$ coupling will be of particular importance in
the following. We point out two key points on
$J_2$-related physics as it has appeared in the literature up to now.
First, the effect of $J_2$ is underestimated in most analytic models
constructed based on the pure iron lattice with only onsite
interactions since the $J_2$ exchange coupling originates mostly from
superexchange processes through As (P) or Se (Te).   Second, in the effective
$\{\tilde{t}\}-J_1-J_2$ model studied before\cite{seo2008}, the
superconducting state is obtained only when the magnetic exchange
coupling strength is of the same order as the hopping parameters (or the
bandwidth) of the model. Therefore, as $t> J$, it requires the
effective bands given by $\{\tilde{t}\}$ be
renormalized. However, the absence of double occupancies in the
standard $t-J$ model is not strictly imposed in such an intermediately
coupled effective model where the bandwidth is assumed to be of similar
order as the interaction scale.

Comparing the predictions from weak coupling and strong coupling, the
122 iron chalcogenides provide an interesting opportunity to address
the difference between the two perspectives. In this paper, we predict
that the $s$-wave pairing symmetry is robust even in extremely
electron-overdoped iron chalcogenides because the AFM $J_2$ is the
main factor for pairing and the $J_1$ is {\it ferromagnetic} (FM), a
conclusion drawn from both neutron scattering
experiments\cite{Lip2011,Ma2009a,Fang2009b} and the magnetic
structure associated with 245 vacancy
ordering\cite{Bao2011a,Fang2011d}. As we will show, the FM $J_1$
significantly reduces the competitiveness of $d$-wave pairing symmetry.
We substantiate this claim by two different methods. First, we
solve the three orbital $\{\tilde{t}\}-J_1-J_2$ model on the mean field
level to show that the $s$-wave pairing is the leading instability
regardless of the change of doping given that $J_2$ is large.  We calculate a full phase diagram  as $J_1$ varies from FM to AFM.
If $J_1$ is AFM,  we obtain a SC state with a mixed s-wave and d-wave
pairing.
%However, the order parameter in this case  is weaker than the case
%when only  $J_2$ being AFM.
Second, we
use the functional renormalization group (FRG) to analyze this trend
obtained by mean field analysis for a 5-band model of the
chalcogenides. We confirm that a dominant AFM $J_2$ generally leads to
robust $s$-wave pairing while an AFM $J_1$ tends to favor $d$-wave pairing
in the electron overdoped region. The competition between $s$-wave
  and $d$-wave weakens the superconducting instability scale. In
addition, it drives the anisotropy feature of the superconducting form
factor as consistently obtained for various weak coupling approaches. Together,
our analysis provides an explanation for the different
behavior of superconductivity in the iron pnictides and iron
chalcogenides in the electron overdoped region since $J_1$ has
opposite signs for these two classes of materials, i. e.  $J_1$ is AFM
in the iron pnictides\cite{Zhaojun2008,Zhaoj2009} and FM in the iron chalcogenides.
Our study suggests that determining the pairing symmetry of the 122 iron chalcogenides can provide important insight regarding whether
 the local AFM exchange couplings are responsible
for the high superconducting transition temperatures.

The paper is organized as follows. In Section~\ref{sec:mfa}, we
present the mean field analysis of the $\tilde{t}-J_1-J_2$ model to
show the differences between the iron pnictide setup $J_1>0$ and the
chalcogenide setup $J_1<0$ in the electron-overdoped regime. This is
followed by FRG studies in Section~\ref{sec:frg} where we mainly
investigate the competition between $s$-wave and $d$-wave in the effective
model, and also analyze the possible effect of an additional pocket at
the $\Gamma$ point of the unfolded BZ which we find to
further increase the robustness of the $s$-wave pairing. The qualitative
trends confirm the results obtained in Sec.~\ref{sec:mfa}. In
Section~\ref{sec:dis} we provide a combined view on the chalcogenides
and point out that the ferromagnetic sign of $J_1$ is important to
explain the robustness of $s$-wave pairing symmetry in these
compounds. Furthermore, we set our work into context of other
approaches to the problem. We conclude in Section~\ref{sec:con} that
electron-overdoped chalcogenides exhibit a robust
$s$-wave pairing phase when the NNN interactions are correctly taken into consideration.

\section{Mean field analysis}
\label{sec:mfa}
\begin{figure}[t]
\includegraphics[width=0.70\linewidth]{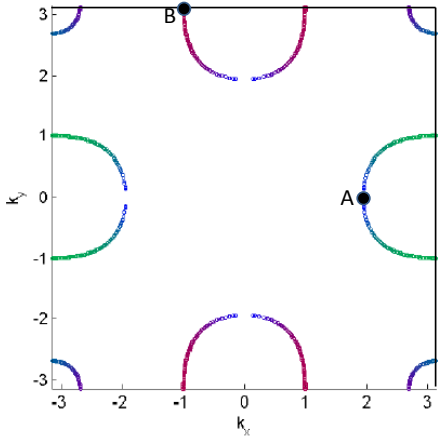}
\caption{(color online) The Fermi surface used to represent the chalcogenides. Colors indicate
the orbital components: Red $d_{xz}$, Green $d_{yz}$, and
Blue=$d_{xy}$. The A and B are auxiliary labels used in Fig.~\ref{fig:MF_orbital}.}
\label{fig:FSandBS}
\end{figure}

We calculate the mean-field
diagram of an effective model for the AFe$_2$Se$_2$ compounds. As the main relevant orbital weight
is given by the $d_{xz}$, $d_{yz}$, and $d_{xy}$ orbital, we employ a
three-orbital kinetic model with $J_1$ and $J_2$ interactions. For
the case of strong electron doping we are interested in, we do
not find qualitative differences when four or five orbital models are
used. For a more thorough discussion of these aspects, refer to Section~\ref{sec:frg}.
The specific kinetic theory we use for the mean-field analysis is a modified
three-band model\cite{Daghofer2010}, given by\bea \hat{T}(k)=\left(
             \begin{array}{ccc}
               T_{11}(k)-\mu & T_{12}(k) & T_{13}(k) \\
               T_{21}(k) & T_{22}(k)-\mu & T_{23}(k) \\
               T_{31}(k) & T_{32}(k) & T_{33}(k)-\mu \\
             \end{array}
           \right), \label{eq:tmat}\eea where
\bea T_{11}(k)&=&2t_2\cos(k_x)+2t_1\cos(k_y)+4t_3\cos(k_x)\cos(k_y),
\nonumber \\
\nonumber T_{22}(k)&=&2t_1\cos(k_x)+2t_2\cos(k_y)+4t_3\cos(k_x)\cos(k_y),\\
\nonumber T_{33}(k)&=&2t_5(\cos(k_x)+\cos(k_y))+4t_6\cos(k_x)\cos(k_y)+\delta,\\
\nonumber T_{12}(k)&=&4t_4\sin(k_x)\sin(k_y),T_{21}(k)=T^\star_{12}(k),\\
\nonumber
T_{13}(k)&=&2it_7\sin(k_x)+4it_8\sin(k_x)\cos(k_y),\\
 T_{23}(k)&=&2it_7\sin(k_y)+4it_8\sin(k_y)\cos(k_x).\eea
The other matrix elements are given by hermiticity. The parameters in the model are taken to be
$t=(0.02,0.06,0.03,-0.01,0.35,0.3,-0.2,0.1)$, $\delta=0.4$, and
$\mu=0.412$. (Throughout the article, energies are given in units of
eV unless stated otherwise). The parameter set chosen gives the Fermi surface shown in Fig.~\ref{fig:FSandBS} with a filling factor of
$4.41$ electrons per site. Aside from a negligibly small electron
pocket at the $M$ point in the unfolded Brillouin zone, the main features are the large electron pockets at $X$
which dictate the physics of the mean-field phase diagram at this electron
doping regime (see Fig.\ref{fig:FSandBS}). In the three band model,
the small electron pocket appears around the M-point in the unfolded
Brillouin zone which may be related to the resonance feature
experimentally discussed for the $\Gamma$ point in the folded
zone. In contrast, the 5-band fit to the chalcogenides we employ in
Section~\ref{sec:frg} suggests small electron pocket features around
the $\Gamma$-point of the unfolded Brillouin zone. Despite this
discrepancy, later we will see that both appearances have a similar
effect and can hence be discussed on the same footing.
The interaction part in our mean field analysis is the pairing energy obtained by decoupling the magnetic exchange couplings\cite{seo2008}, which  can be written as
\bea
\hat{V}&=&-\sum_{\alpha,r}(J_1b^\dag_{\alpha,r,r+x}b_{\alpha,r,r+x}+J_1b^\dag_{\alpha,r,r+y}b_{\alpha,r,r+y}
\\\nonumber&&+J_2b^\dag_{\alpha,r,r+x+y}b_{\alpha,r,r+x+y}+J_2b^\dag_{\alpha,r,r+x-y}b_{\alpha,r,r+x-y})\eea

where
$b_{\alpha,r,r'}=c_{\alpha,r,\uparrow}c_{\alpha,r',\downarrow}-c_{\alpha,r,\downarrow}c_{\alpha,r',\uparrow}$
represent singlet pairing operators between the $r,r'$ sites.

Before we perform the
self-consistent mean field calculation, we define the pairing order parameters as follows:
in real space, the pairings on two NN bonds and two NNN bonds are
 represented by $\Delta^{\alpha}_{x}=J_1<b_{\alpha,r,r+x}>$, $\Delta^\alpha_y=J_1<b_{\alpha,r,r+y}>$,
 $\Delta^\alpha_{x+y}=J_2<b_{\alpha,r,r+x+y}>$ and $\Delta^\alpha_{x-y}=J_2<b_{\alpha,r,r+x+y}>$, where $\alpha$
 denotes the orbital index and $x,y$ are the two unit lattice
 vectors. We only consider intra-orbital pairing and ignore
 inter-orbital pairing which is very small as shown in previous
 calculations\cite{seo2008}. Considering the $C_4$  symmetry of the
 lattice,  we can classify the pairing symmetries according to  the
 one-dimensional irreducible  representations of the $C_4$
 symmetry. Since the pairing is a spin singlet, we can classify them
 as follows: an order parameter is of A-type (B-type) if it is even
 (odd) under a 90-degree rotation.
This classification leads to six candidate pairings with
A-symmetry and another six candidates with B-symmetry as the SC pairings include NN from $J_1$ and
NNN from $J_2$ bonds, which manifests as
the A-type
symmetry\bea\Delta^A_{NN,s}&=&(\Delta^{xz}_x+\Delta^{xz}_y+\Delta^{yz}_x+\Delta^{yz}_y)/4,
\nonumber\\
\nonumber\Delta^A_{NN,d}&=&(\Delta^{xz}_x-\Delta^{xz}_y-\Delta^{yz}_x+\Delta^{yz}_y)/4,\\
\nonumber\Delta^A_{NNN,s}&=&(\Delta^{xz}_{x+y}+\Delta^{xz}_{x-y}+\Delta^{yz}_{x+y}+\Delta^{yz}_{x-y})/4,\\
\nonumber\Delta^A_{NNN,d}&=&(\Delta^{xz}_{x-y}-\Delta^{xz}_{x+y}+\Delta^{yz}_{x+y}-\Delta^{yz}_{x-y})/4,\\
\nonumber\Delta^{xy}_{NN,s}&=&(\Delta^{xy}_{x}+\Delta^{xy}_y)/2,\\
\Delta^{xy}_{NNN,s}&=&(\Delta^{xy}_{x+y}+\Delta^{xy}_{x-y})/2.\eea
and the B-type
symmetry \bea\Delta^B_{NN,s}&=&(\Delta^{xz}_x+\Delta^{xz}_y-\Delta^{yz}_x-\Delta^{yz}_y)/4,\nonumber
\\
\nonumber\Delta^B_{NN,d}&=&(\Delta^{xz}_x-\Delta^{xz}_y+\Delta^{yz}_x-\Delta^{yz}_y)/4,\\
\nonumber\Delta^B_{NNN,s}&=&(\Delta^{xz}_{x+y}+\Delta^{xz}_{x-y}-\Delta^{yz}_{x+y}-\Delta^{yz}_{x-y})/4,\\
\nonumber\Delta^B_{NNN,d}&=&(\Delta^{xz}_{x-y}-\Delta^{xz}_{x+y}-\Delta^{yz}_{x+y}+\Delta^{yz}_{x-y})/4,\\
\nonumber\Delta^{xy}_{NN,d}&=&(\Delta^{xy}_{x}-\Delta^{xy}_y)/2,\\
\Delta^{xy}_{NNN,d}&=&(\Delta^{xy}_{x-y}-\Delta^{xy}_{x+y})/2.\eea
In reciprocal lattice space, the mean-field Hamiltonian  is given by \bea
\label{eq:MF}\hat{H}=\sum_k\left(
                                                                       \begin{array}{cc}
                                                                         \hat{T}(k) & \hat\Delta(k) \\
                                                                         \hat\Delta^\dag(k) & -\hat{T}^\star(-k) \\
                                                                       \end{array}
                                                                     \right),\eea where\bea\hat\Delta(k)=\left(
                                                                                                           \begin{array}{ccc}
                                                                                                             \Delta_{11}(k) & 0 & 0 \\
                                                                                                              0& \Delta_{22}(k) & 0 \\
                                                                                                              0& 0 & \Delta_{33}(k) \\
                                                                                                           \end{array}
                                                                                                         \right),\eea and
                                                                                                         \begin{widetext}
\bea\Delta_{11}(k)&=&(\Delta_{NN,s}^A+\Delta^B_{NN,s})(\cos(k_x)+\cos(k_y))+(\Delta_{NN,d}^A+\Delta^B_{NN,d})(\cos(k_x)-\cos(k_y))\nonumber
\\
\nonumber&&+2(\Delta_{NNN,s}^A+\Delta^B_{NNN,s})\cos(k_x)\cos(k_y)+2(\Delta_{NNN,d}^A+\Delta^B_{NNN,d})\sin(k_x)\sin(k_y),\\
\nonumber\Delta_{22}(k)&=&(\Delta_{NN,s}^A-\Delta^B_{NN,s})(\cos(k_x)+\cos(k_y))+(\Delta_{NN,d}^B-\Delta^A_{NN,d})(\cos(k_x)-\cos(k_y))\\
\nonumber&&+2(\Delta_{NNN,s}^A-\Delta^B_{NNN,s})\cos(k_x)\cos(k_y)+2(\Delta_{NNN,d}^B-\Delta^A_{NNN,d})\sin(k_x)\sin(k_y),\\
\nonumber\Delta_{33}(k)&=&\Delta^{xy}_{NN,s}(\cos(k_x)+\cos(k_y))+\Delta^{xy}_{NN,d}(\cos(k_x)-\cos(k_y))+2\Delta^{xy}_{NNN,s}\cos(k_x)\cos(k_y)\nonumber\\
&&+2\Delta^{xy}_{NNN,d}\sin(k_x)\sin(k_y).\eea
\end{widetext} We emphasize that in above definitions the
  symbols $s$ and $d$ merely represent the geometric factor of pairing
  in $k$-space, and do {\it not} correspond to whether pairing is even or odd under a 90-degree rotation in a multi-orbital system.
In general, there are more than one self-consistent set of
$\{\Delta\}$'s as self-consistent meanfield solutions. The free
energies in each solution hence have to be compared
to find the solution with the lowest free energy.

\begin{figure}[t]
\includegraphics[width=6cm]{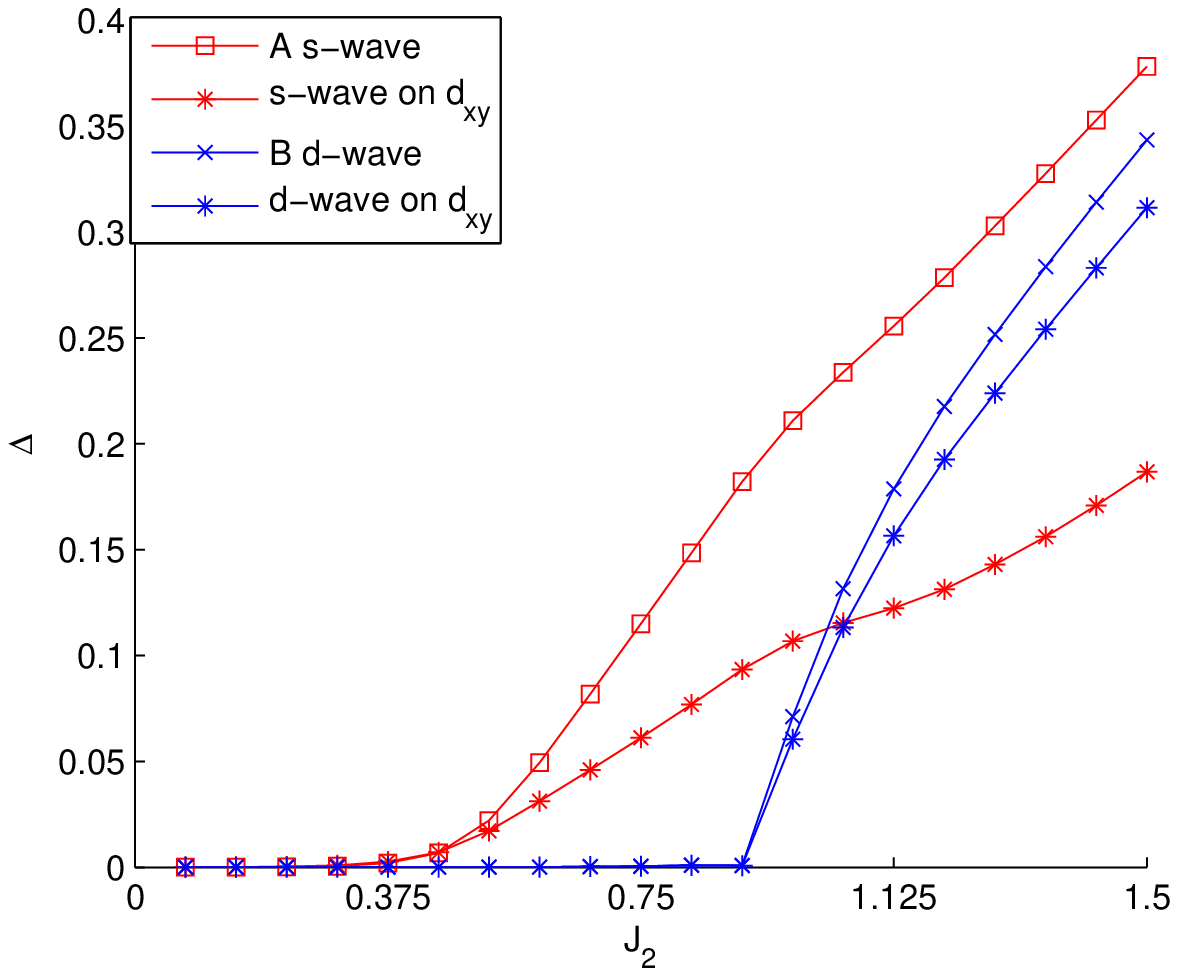}
\includegraphics[width=6cm]{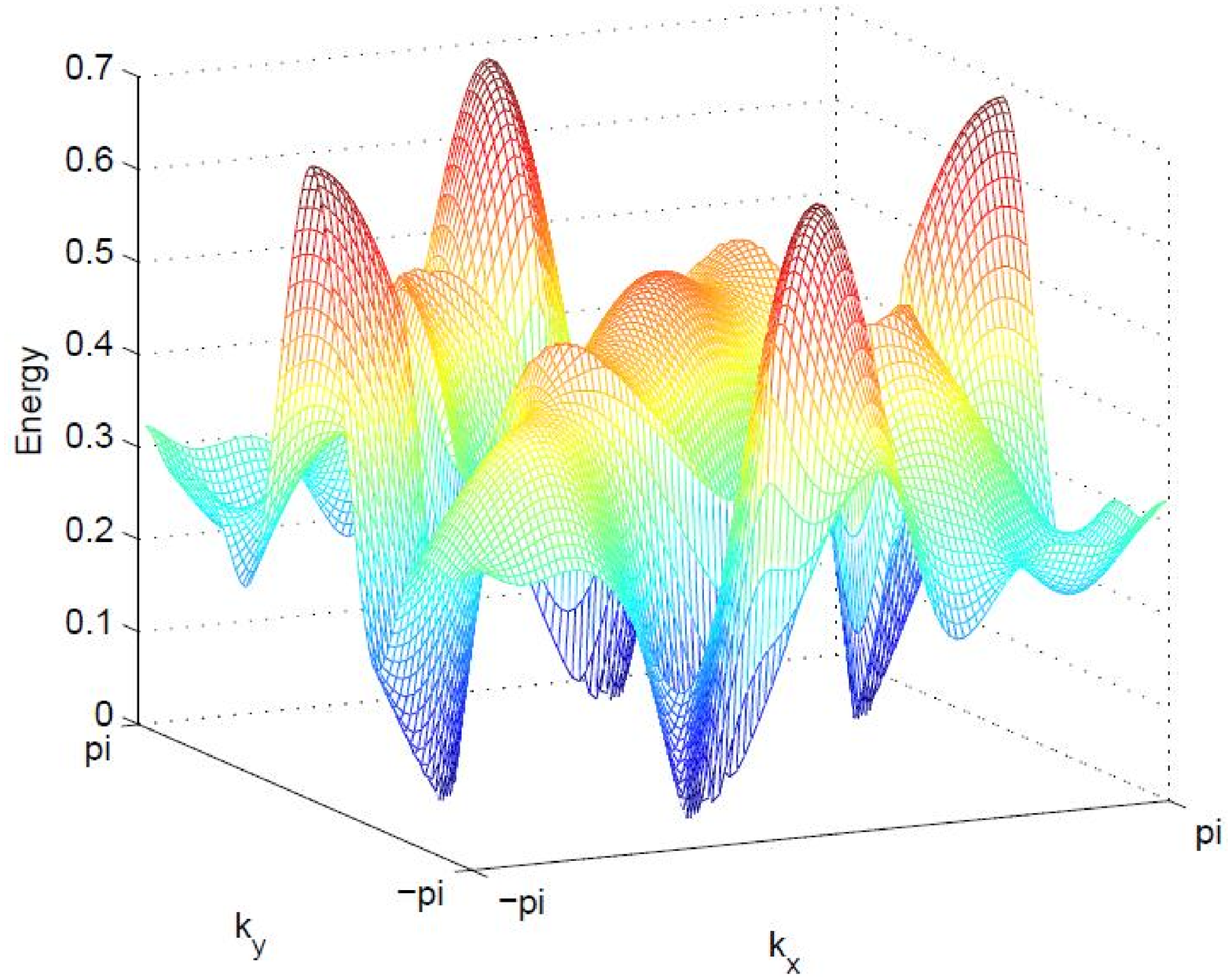}
\includegraphics[width=6cm]{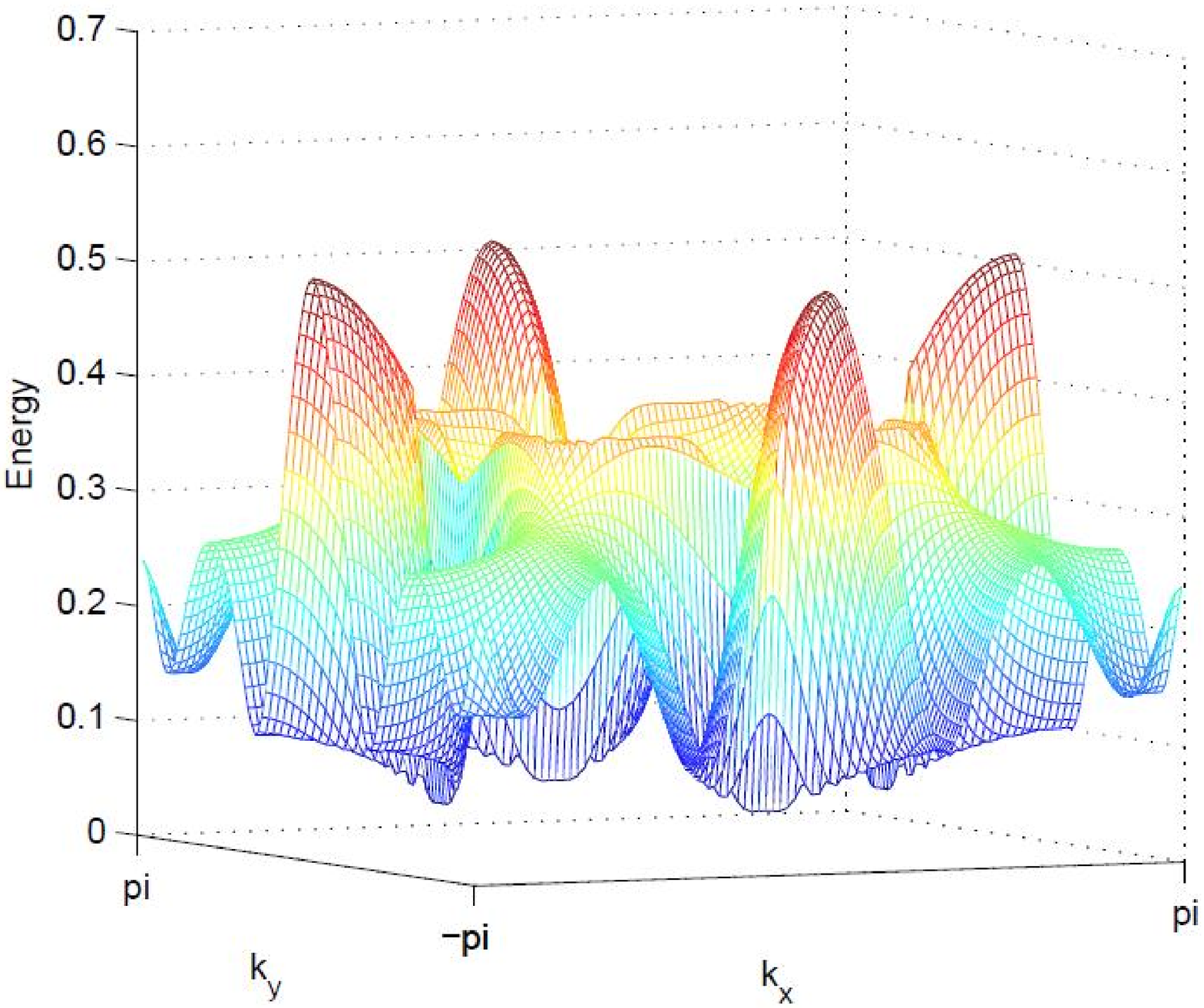}
\caption{(color online) The mean field phase diagram along $J_1=0$ in
  the parameter space (up); the quasiparticle spectrum at $J_2=1$
  (middle) and $J_2=0.75$ (bottom). At $J_2=1$, one can see that the quasiparticle spectrum explicitly breaks $C_4$ symmetry because of the mixing of A- and B-type pairing symmetries.\label{fig:J2scan}}
\end{figure}

First consider pure NNN-pairings stemming from $J_2$ (this is a
reasonable limit to start with since $J_1$ in
FeTe(Se) has been shown to be ferromagnetic, thus not
contributing to pairing in the singlet pairing channel). $J_2$ is
increased from zero to $J_2=1.5$  while the band width is
$W\sim4$. The robust superconductivity solution with purely A-type
$s$-wave pairing is obtained when  $J_2$ is larger than $0.4$. This is
to say the pairing remains the same as in iron-pnictides with the
geometric factor $\cos(k_x)\cos(k_y)$\cite{seo2008}. The Bogoliubov
particle spectrum is completely gapped in this state. When $J_2$
becomes larger than 1, the ground state is a mixture of A- and B-type
pairings. The nonzero B-type pairings all have the geometric factor
$\sin(k_x)\sin(k_y)$ (see the phase diagram show in Fig.~\ref{fig:J2scan}). In the coexistence phase, the quasiparticle spectrum shows
nearly gapless features at several points, and moreover, the
dispersion explicitly breaks $C_4$ rotation symmetry (see
Fig.~\ref{fig:J2scan} displaying the quasiparticle spectrum of the lowest
branch).

\begin{figure}[t]
\includegraphics[width=6cm]{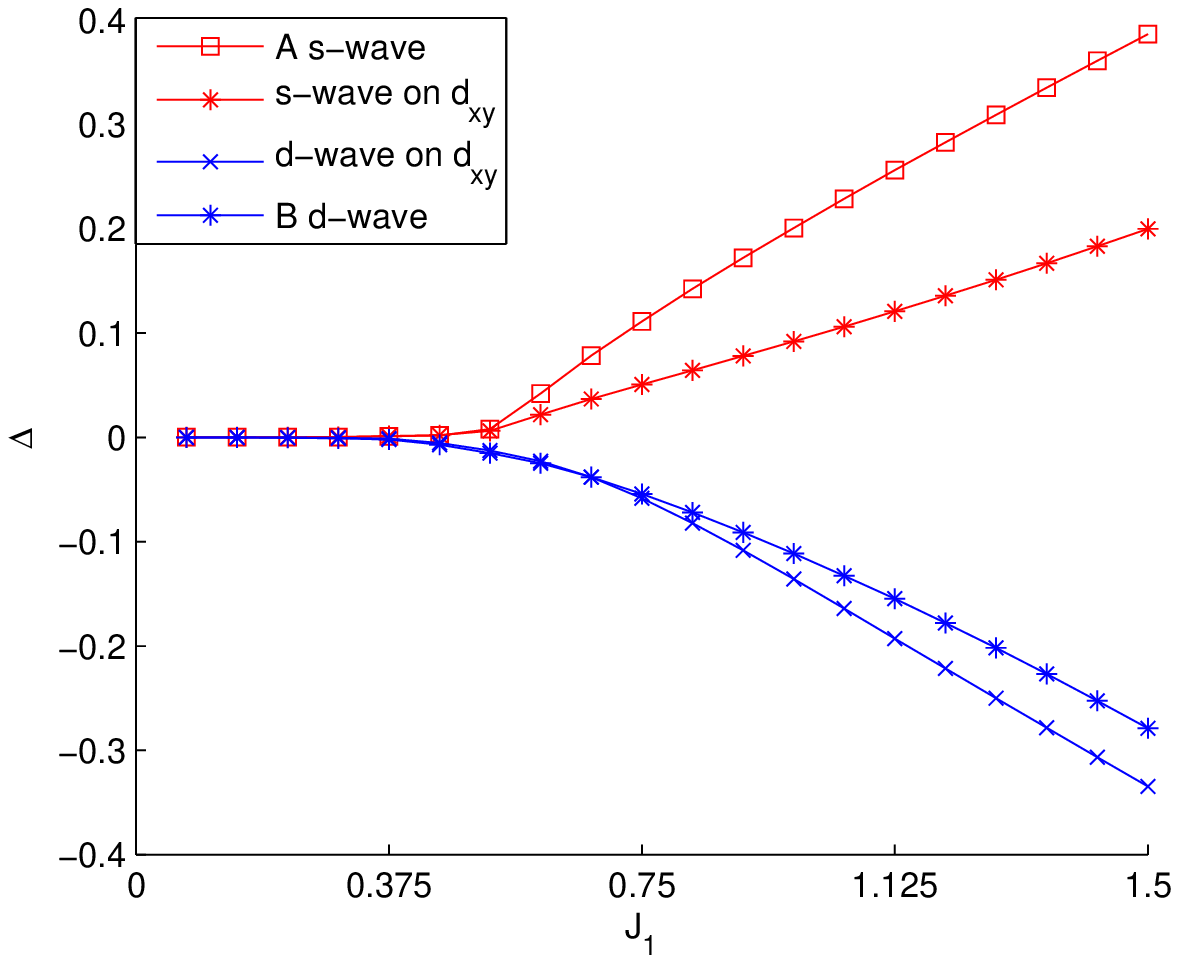}
\includegraphics[width=6cm]{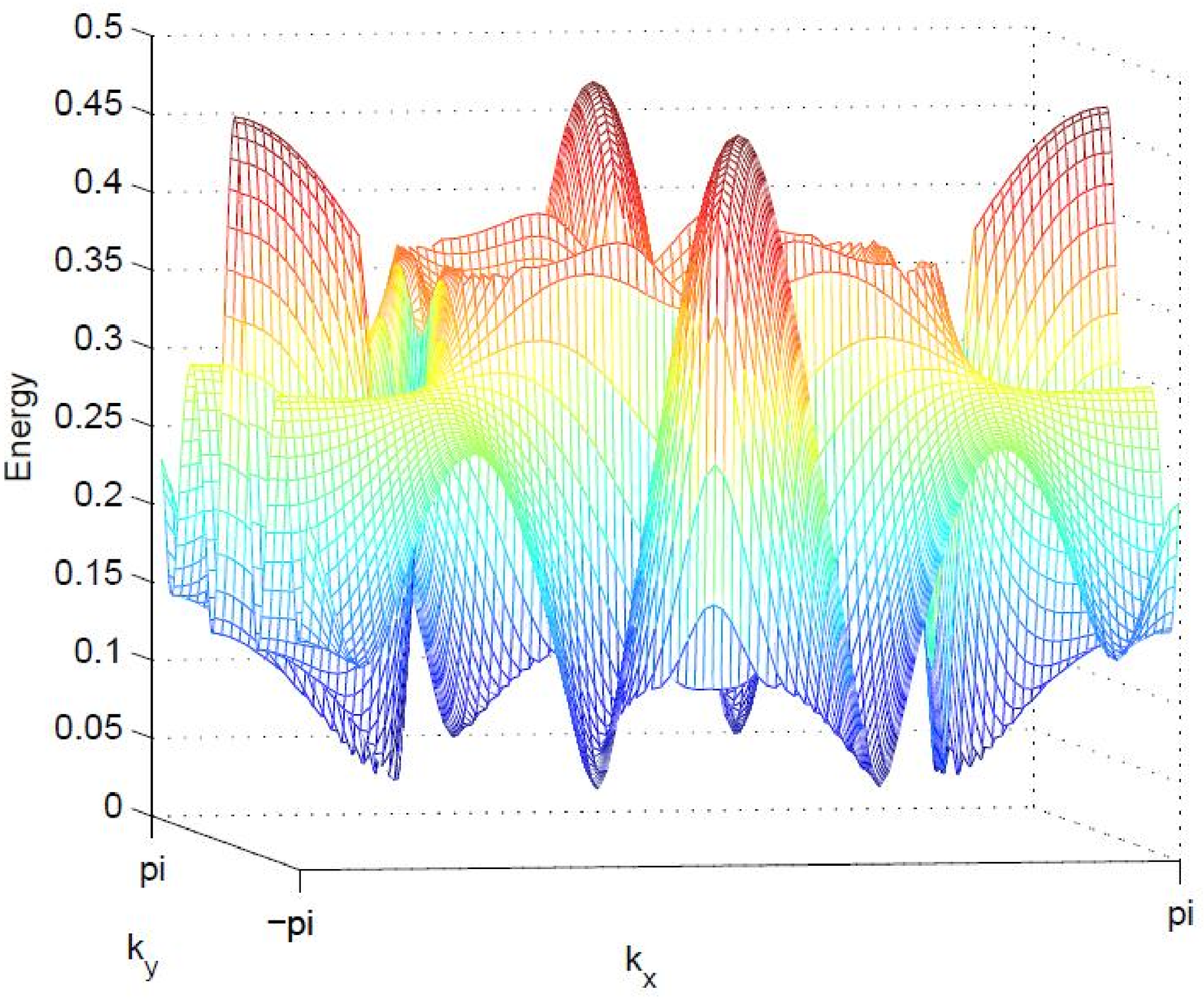}
\caption{(color online) The mean field phase diagram along $J_2=0$ in the parameter
space (up); the quasiparticle spectrum at $J_1=0.75$
(bottom). At $J_1=0.75$ one can see that the quasiparticle spectrum explicitly breaks the $C_4$ symmetry because of the mixing of A- and B-type pairing symmetries.\label{fig:J1scan}}
\end{figure}
Second, we  study the phase diagram when only (antiferromagnetic) $J_1$ is
present. In this case, only NN pairings are nonzero and there are
six SC gaps. We increase $J_1$ from $J_1=0$ to $J_1=1.5$
where the band width is $W\sim4$. The SC order becomes non-zero from
$J_1=0.4$ on. However,  in this case,  the
B-type SC order arises slightly earlier than A-type SC order. The
ground state is always a mixture of A- and B-type pairings. The two
leading orders are A-type $s$-wave and B-type $d$-wave in $xz,yz$
orbitals while the sub-leading ones are $s$- and $d$-waves in the $xy$
orbitals (Fig.~\ref{fig:J1scan}). Due to
strong mixing of A- and B-type pairings,  the quasiparticle spectrum
is very anisotropic. It is, however, still nodeless, in contrast to a pure
$s$-wave pairing $\cos(k_x)+\cos(k_y)$ where there are nodes\cite{Thomale2010asp} on the
electron pockets (see Fig.~\ref{fig:J1scan}).

\begin{figure}[t]
\includegraphics[width=8cm]{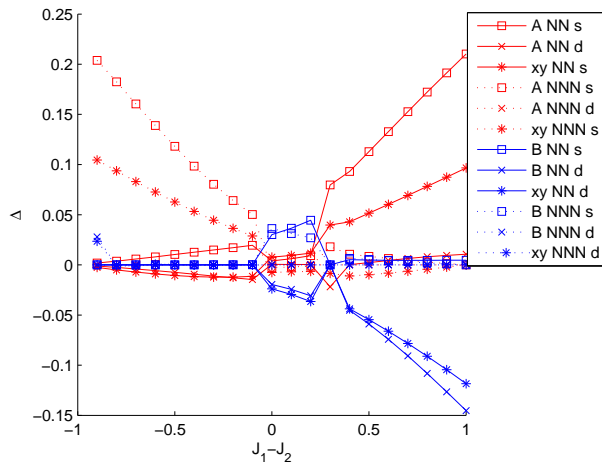}
\caption{(color online) The mean field phase diagram with $J_1+J_2=1$ in the parameter space.\label{fig:J1-J2scan}}
\end{figure}
\begin{figure}[t]
\includegraphics[width=8cm]{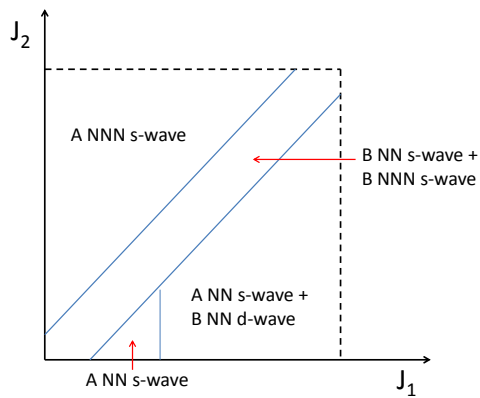}
\caption{(color online) A schematic phase diagram for the model~\eqref{eq:MF} within $0<J_1,J_2<1$.\label{fig:phase_diagram}}
\end{figure}

Finally, for $J_1$ and $J_2$ antiferromagnetic, we fix
$J_1+J_2=1$ and change $J_1-J_2$ as a parameter. We observe
that NNN pairings dominate for
$J_1-J_2<-0.1$ and NN pairings dominate for $J_1-J_2>0.2$ (Fig.~\ref{fig:J1-J2scan}). In the
intermediate range, there is only weak B-type pairing. A
schematic phase diagram  within the range $0<J_1,J_2<1$ is shown
in Fig.~\ref{fig:phase_diagram}.

\begin{figure}[t]
\includegraphics[width=8cm]{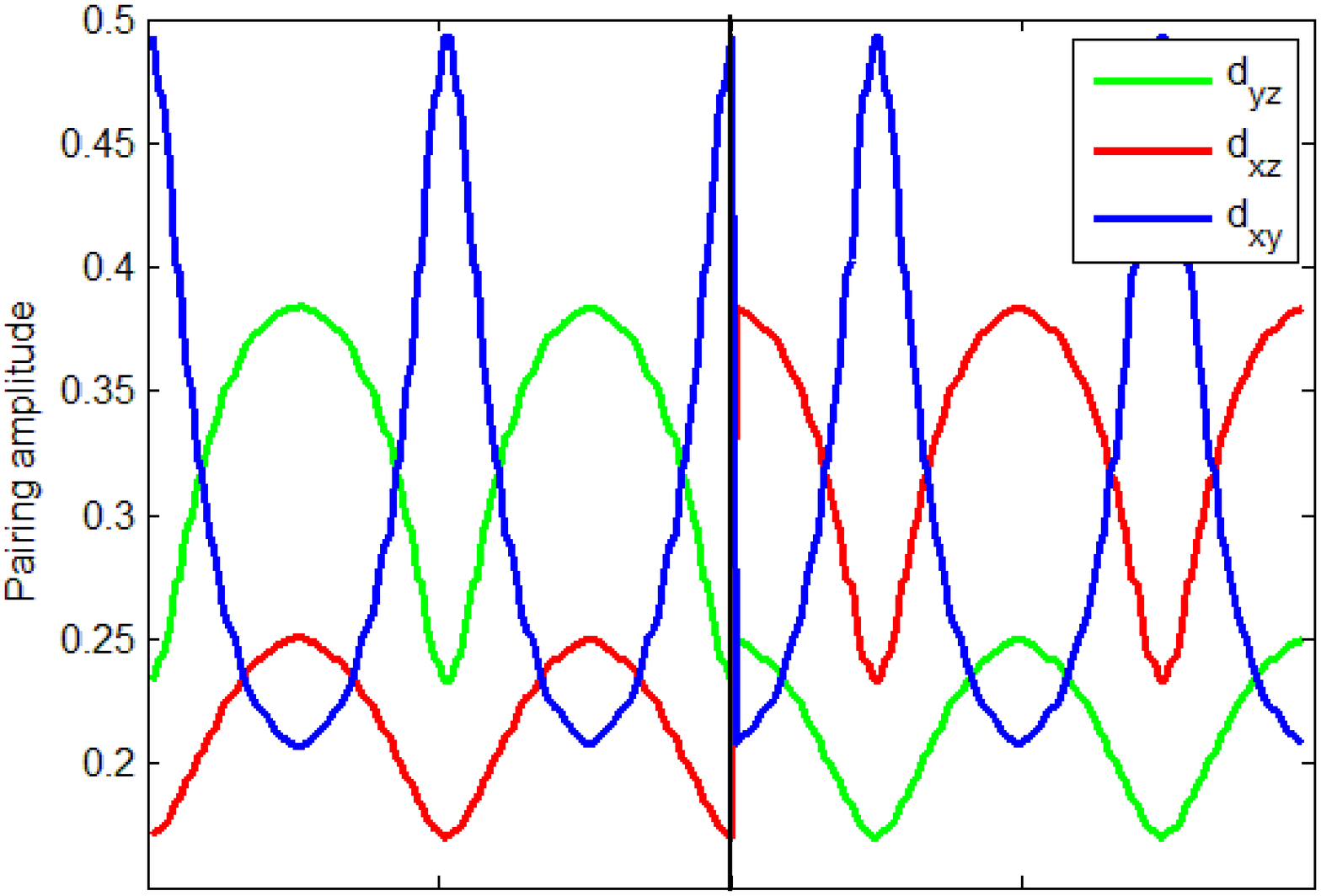}
\includegraphics[width=8cm]{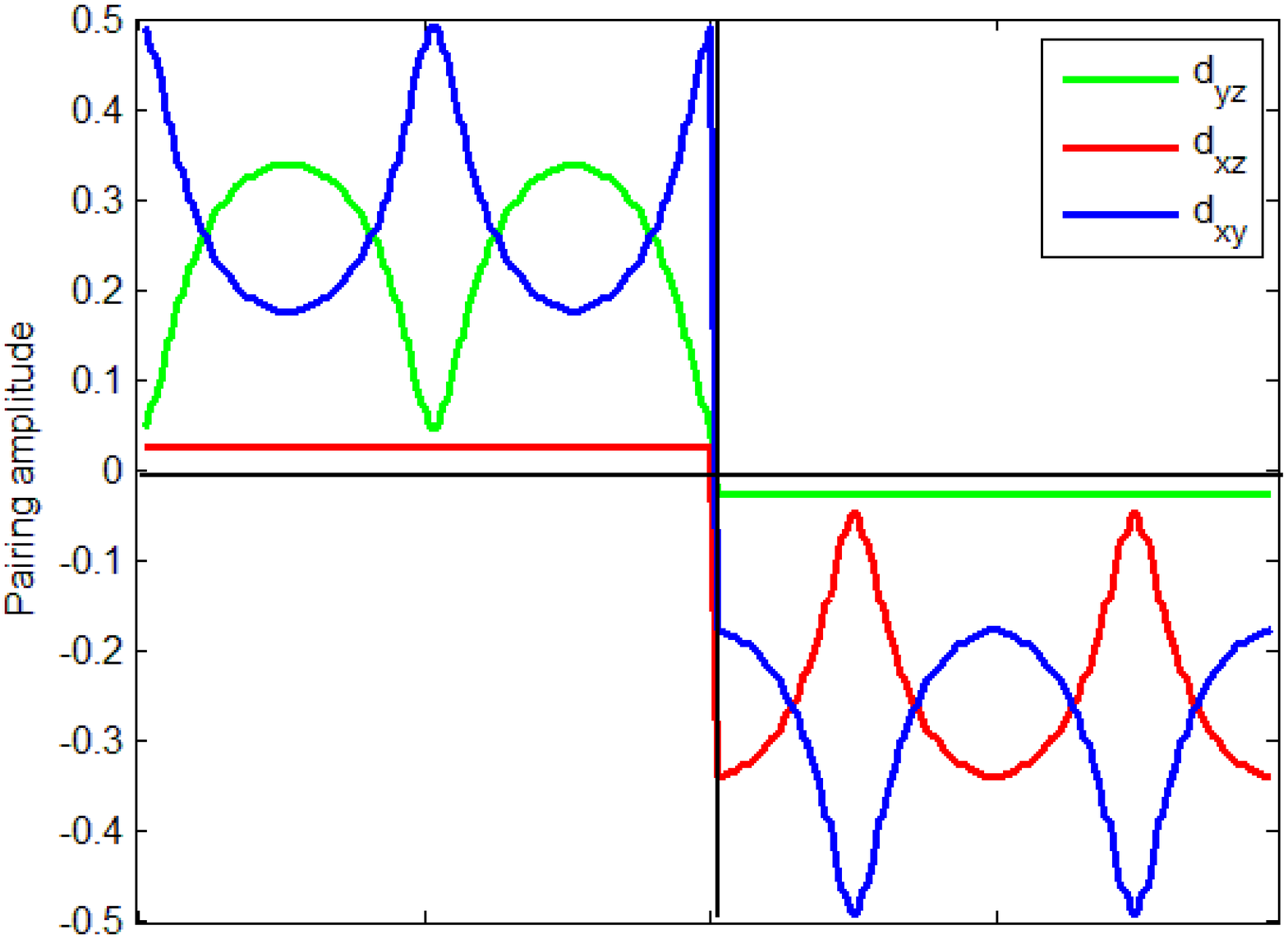}
\caption{(color online) The orbital resolved pairing amplitude on the
  FS for a typical $s$-wave ($d$-wave) pairing state in the upper
  (lower) panel, calculated within meanfield approximation. The
  interaction parameters are $J_1=0$, $J_2=0.8$ for the upper panel
  and $J_1=0.5$, $J_2=0$ for the lower panel. In the left half of
  these figures, the k-point traces the electron pocket around $X$-point
  counterclockwise from point A in Fig.~\ref{fig:FSandBS} and in the
  right half, it traces the electron pocket around the $Y$-point counterclockwise from point B in Fig.~\ref{fig:FSandBS}.\label{fig:MF_orbital}}

\end{figure}

In the whole parameter region of $(J_1,J_2)$, the SC order
  parameters always have the same sign for all three orbitals. This
  can be seen in Fig.~\ref{fig:MF_orbital} where the orbital resolved
  pairing amplitude is shown along electron pockets around $X$. This
  result is essentially consistent with the FRG result shown in Sec.~\ref{sec:frg} (Fig.\ref{fig:orbital-two}). It is, however, different from
  what one would expect from the very strong coupling limit: There,
  the strong inter-orbital repulsion favors different signs of pairing
  for the $d_{xy}$ orbital and the $d_{xz/yz}$ orbitals\cite{Lu2010}. Some quantitative differences between Fig.~\ref{fig:MF_orbital} and Fig.~\ref{fig:orbital-two} may be explained as the incompleteness of a three-orbital model and the fact that the meanfield pairing is not constrained to the FS. In Fig.~\ref{fig:MF_orbital} we also see that the orbital resolved pairing amplitude is highly anisotropic: This is a natural reflection of different orbital composition on different parts of the Fermi surface.

Following the Fermi surface topology in Fig.~\ref{fig:FSandBS}, these
mean-field results have been obtained in the case where a small
electron pocket was still present at the $M$-point.
For completeness of the analysis, we also adapted the parameters such
that the electron pocket around the $M$ point vanishes, leaving two
pockets around $X$. Without the $M$ pocket, we see that $s$-wave
pairings are less favored than before, as its geometric factor is
$\cos(k_x)\cos(k_y)$ or $\cos(k_x)+\cos(k_y)$, both being maximized
around $M$. With the two-pocket FS, taking $J_1=0$ and increasing
$J_2>0$, B-type pairings blend in at smaller $J_2$ than shown in
Fig.\ref{fig:J2scan}; taking $J_2=0$ and increasing $J_1>0$, A-type
pairings appear at slightly larger $J_1$ than shown in
Fig.~\ref{fig:J1scan}. The main features still remain unchanged. These
trends are in accordance with the FRG studies in the following section.

\section{FRG analysis}
\label{sec:frg}

To substantiate the mean field results above, we employ functional
renormalization group (FRG)\cite{zanchi-00prb13609,halboth-00prb7364,honerkamp-01prb035109} to further
investigate the pairing symmetry of the $\tilde{t}$-$J_1$-$J_2$
model. As an unbiased resummation scheme of all
channels, the
FRG has been extended and amply employed to the multi-band case of iron
pnictides. More details can be found in
Refs.~\onlinecite{Wang2008a, platt-09njp055058, Thomale2009, Thomale2010asp}. The conventional starting point for the FRG are bare Hubbard-type
interactions which develop different Fermi surface instabilities as
higher momenta are integrated out when the cutoff of the theory flows to
the Fermi surface. To address the special situation found in the
chalcogenides where the Fe-Se coupling is strong, not only
local, but also further neighbor interaction terms
would have to be taken into account:
in our FRG setup, the onsite Hubbard-type interactions of the same type as
used in the study of pnictides triggers no instability at reasonable
critical scales. This suggests already at this stage that the
chalcogenides may necessitate a perspective beyond pure weak coupling.
In addition, the total parameter
space of bare interactions is large and constrained RPA
parameters are not yet available for this class of materials.
For the purpose of our study, we hence constrain ourselves to the
$\tilde{t}$-$J_1$-$J_2$ model from the outset. This implies that the
pairing interaction is already attractive on the bare level, and a
development of an SC instability is expected as the physics is
dominated by the pairing channel.
%In turn, we expect the
%competition between different parquet channels to be less relevant and
%a good correspondence of the results obtained here and other approaches
%such as RPA in the pairing channel.
Still, we can employ FRG to investigate the properties and competition of different SC pairing
symmetries for the chalcogenides for different $(J_1,J_2)$ regimes.

Within FRG, we consider general $J_1$-$J_2$ interactions which are not limited to the spins
in the same orbital:
\begin{eqnarray*}
H&=&J_1\sum_{\langle i,j\rangle}\sum_{a,b}
(\mathbf{S}_{ia}\cdot\mathbf{S}_{jb}-\frac{1}{4}n_{ia}n_{jb})\\
&+&J_2\sum_{\langle\langle i,j\rangle\rangle}\sum_{a,b}
(\mathbf{S}_{ia}\cdot\mathbf{S}_{jb}-\frac{1}{4}n_{ia}n_{jb}).
\end{eqnarray*}
The kinetic theory will differ in the various cases studied below. For
all cases, we will study the full 5-band model incorporating all Fe
$d$ orbitals. Concerning
the discretization of the BZ, the RG calculations were
performed with 8 patches per pocket, and a $10_\text{radius}\times
3_\text{angle}$ mesh on each patch. (This moderate resolution is
convenient to scan wide ranges of the interaction parameter space; we
checked that increasing the BZ resolution did not qualitatively change
our findings.) The output of the
RG calculation is the four-point vertex on the Fermi surfaces: $V_{\Lambda}(\bs{k}_1,n_1;\bs{k}_2,n_2;\bs{k}_3,n_3;\bs{k}_4,n_4)c_{\bs{k}_4n_4s}^{\dagger}c_{\bs{k}_3n_3\bar{s}}^{\dagger}c_{\bs{k}_2n_2s}^{\phantom{\dagger}}c_{\bs{k}_1n_1\bar{s}}^{\phantom{\dagger}},$
%\end{equation*}
where the flow parameter is the IR cutoff $\Lambda$ approaching the
Fermi surface, and with $\bs{k}_{1}$ to $\bs{k}_{4}$ the incoming and
outgoing momenta. We only find singlet pairing to be relevant for the
scenarios studied by us:
$\sum_{\bs{k},\bs{p}}V_{\Lambda}(\bs{k},\bs{p})
[\hat{O}^{\dagger}_{\bs{k}}
\hat{O}^{\phantom{\dagger}}_{\bs{p}}]$, where
$\hat{O}^{\text{SC}}_{\bs{k}}=c_{\bs{k},\uparrow}c_{-\bs{k},\downarrow}$.
We decompose
the pairing channel into eigenmodes,
\begin{equation}
V^{\text{SC}}_{\Lambda} (\bs{k},-\bs{k},\bs{p})= \sum_i c_i^{\text{SC}}(\Lambda) f^{\text{SC},i}(\bs{k})^* f^{\text{SC},i}(\bs{p}),
\label{decomp}
\end{equation}
and obtain the band-resolved form factors of the leading and
subleading SC eigenmode (i.e. largest two negative eigenvalues). This
way we are able to discuss the interplay of $d$-wave and $s$-wave as well
as the degree of form factor anisotropy for a given setting of $(J_1,
J_2)$. Comparing divergence scales $\Lambda_c$ gives us the
possibility to investigate the relative change of $T_c$ as a function
of $(J_1, J_2)$. Furthermore, we also investigate the orbital-resolved
pairing modes\cite{Thomale2010asp} by decomposing the orbital four
point vertex
\begin{eqnarray}
V^{\text{orb}}_{c,d \rightarrow a,b}&=& \sum_{n_1,\ldots,n_4 = 1}^5 \Big{\{}V_{\Lambda}(\bs{k}_1,n_1;\bs{k}_2,n_2;\bs{k}_3,n_3;\bs{k}_4,n_4)\nonumber \\
&&\times u^*_{an_1}(\bs{k}_1)u^*_{bn_2}(\bs{k}_2)u_{cn_3}(\bs{k}_3)u_{dn_4}(\bs{k}_4)\Big{\}},
\label{orbital}
\end{eqnarray}
where the $u$'s denote the different orbital components of the band
vectors. By investigating the intraorbital SC pairing modes in
\eqref{orbital}, we make contact to the findings from the previous mean field analysis.

\subsection{Two-pocket scenario}
\label{sec:tps}

\begin{figure}[t]
\includegraphics{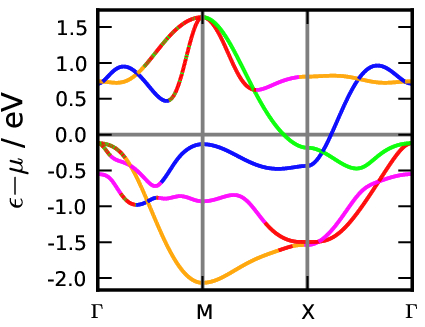}%
\includegraphics{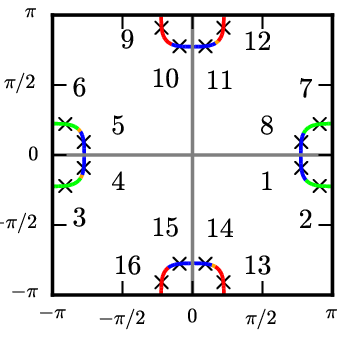}%
\caption{(color online) The spectrum and the Fermi surfaces of the
band structure proposed by Maier et al.~\cite{Maier2011}, colored according to
the dominant orbital content. The color code is red $d_{xz}$, green $d_{yz}$, blue
$d_{xy}$, orange $d_{x^2-y^2}$ and magenta $d_{3z^2-r^2}$. The numbered
crosses show the center of Fermi surface patches used in the FRG
calculations.}
\label{bandmaier}
\end{figure}

We start by studying the 5-band model suggested before by Maier et
al.\cite{Maier2011}. There are only two electron pockets at the $X$ point
of the unfolded Brillouin zone closely resembling the Fermi surface
topology and orbital content employed for our mean-field analysis (Fig.~\ref{bandmaier}).

\begin{figure}[t]
\includegraphics{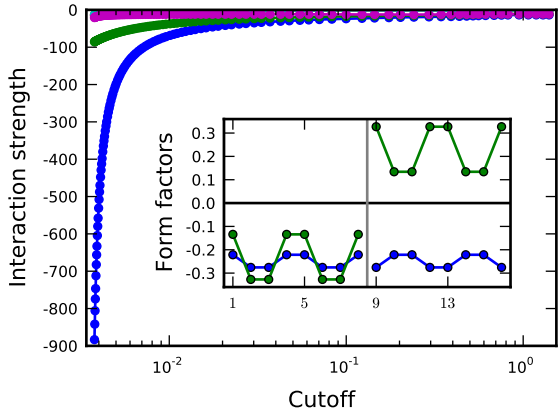}
\caption{\label{fig:j1j2-typical1} (color online) Typical RG flows and the
superconducting gaps associated with the Fermi surfaces for the two-pocket
scenario with $(J_1,J_2)=(0.1,0.5)$~eV. Leading form
factor is denoted in blue, sub-leading form factor in green.}
\end{figure}

\begin{figure}[t]
\includegraphics{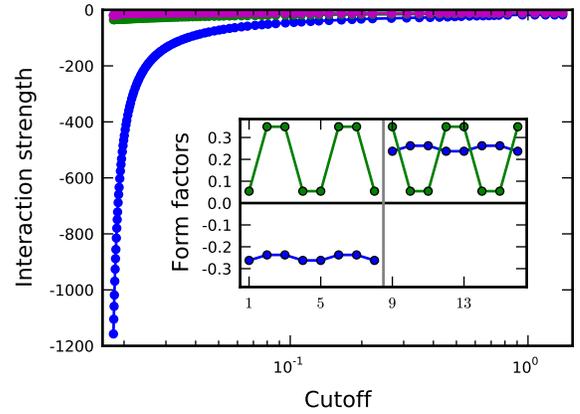}
\caption{\label{fig:j1j2-typical2} (color online) Typical RG flows and the
superconducting gaps associated with the Fermi surfaces for the two-pocket
scenario with $(J_1,J_2)=(0.5,0.1)$~eV. Leading form
factor is denoted in blue, sub-leading form factor in green.}
\end{figure}

The RG flow
and the form factors of the leading diverging channels are shown in
Fig.~\ref{fig:j1j2-typical1} for dominant $J_2$ and in
Fig.~\ref{fig:j1j2-typical2} for dominant $J_1$. As stated before, the pairing
interaction is already present at the bare level in the model so
that we achieve comparably fast instabilities as the cutoff is
flowing towards the Fermi surface. As found in Ref.~\onlinecite{seo2008}, the
dominant $J_2$ scenario exhibits a leading $s$-wave $\cos k_x \cos k_y$ form
factor which causes the same sign on both electron pockets (blue dots
in Fig.~\ref{fig:j1j2-typical1}). The subleading form factor is found
to be of $d$-wave $\cos k_x-\cos k_y$ type, changing sign
from one electron pocket to the other.
The inverse situation is found for dominant $J_1$. As shown in
Fig.~\ref{fig:j1j2-typical2}, the  $d$-wave $\cos k_x-\cos k_y$ form
factor establishes the leading instability. As before, the form factor
does not cross zero related to nodeless SC for this parameter setting.

With only pairing information available on the limited number of sampling points
along FS, it is impossible to obtain as
in the mean-field analysis the superconducting gap in the whole BZ.
For illustration, a mixture of a small
A-type NNN d-wave pairing and large A-type NNN s-wave pairing is
indistinguishable from a pure A-type NNN s-wave pairing; a mixed state
of a small B-type NN $s$-wave pairing plus a large B-type NN $d$-wave
pairing, and a state with pure B-type NN $d$-wave pairing show little
difference if one compares the gap on a few points along the Fermi
surfaces. For this reason, the symbol $s_{x^2y^2}$ used in this
section refers to a pairing consisting of a large A-type NNN $s$-wave
pairing and possible small components of A-type NNN $d$-wave pairing
or A-type NN $s$/$d$-wave pairing. In turn, the symbol $d_{x^2-y^2}$
refers to a pairing made up with a large B-type NN $d$-wave pairing
and possible small components of B-type NN $s$-wave pairing or B-type
NNN $s$/$d$-wave pairing.


\begin{figure}[t]
\centering
\includegraphics{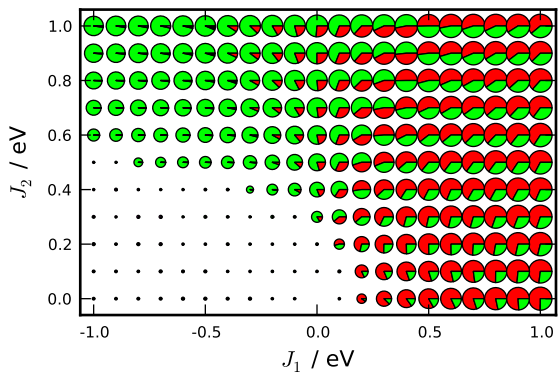}%
\caption{\label{fig:phase-two} (color online) The phase diagram of the two-pocket
model. Each pie shows the relative strengths of the two leading pairing channels, with the radius proportional to
$[8+\log_{10}(\Lambda_c/\text{eV})]^2$. The color code for pairing symmetries
is green $s_{x^2y^2}$ and red $d_{x^2-y^2}$.}
\end{figure}

We have scanned a large range of $(J_1,J_2)$. For each setup we
have obtained $\Lambda_c$ as well as the ratio of the instability
eigenvalues between $s$-wave and $d$-wave in the pairing channel (encoded
by the two-color circles shown in Fig.~\ref{fig:phase-two}). The FRG result is qualitatively
consistent with the mean field analysis. In the antiferromagnetic
sector, the $s$-wave wins for dominant $J_2$ while
the $d$-wave wins wins for dominant $J_1$. For ferromagnetic $J_1$
corresponding to the situation in chalcogenides, we find a robust
preference of $s$-wave pairing. {The anisotropy of the s-wave gap around the pockets in the FRG calculation is also qualitatively consistent with the meanfield result. The gap on the Fermi surfaces with $d_{xy}$ orbital character is smaller than the gap on those with $d_{xz,yz}$ orbital character. }

\begin{figure}[t]
\includegraphics{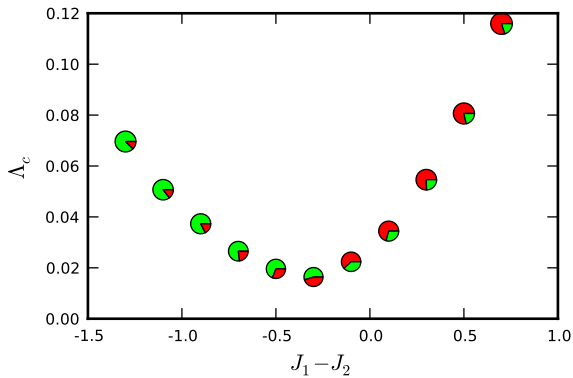}%
\caption{\label{fig:svsd-two} (color online) The variation of the critical scale $\Lambda_c$
along a line through parameter space which interpolates between $s$
and $d$ wave. A minimum is visible for comparable ordering tendency in
$s$-wave and $d$-wave.  See the caption of Fig.~\ref{fig:phase-two} for
more details on the pie charts.}
\end{figure}

The predictions from the mean field analysis are further confirmed for
the mixed
phase regime where $s$-wave and $d$-wave coexist in the mean field
solution. In FRG, one of these instabilities
will always be slightly preferred; still, when both
instabilities diverge in very close proximity to each other, this
regime behaves similarly to the coexistence phase. For illustration,
in Fig.~\ref{fig:svsd-two} we have plotted the dependence of
$\Lambda_c$ on $J_1-J_2$, with $J_1+J_2$ fixed to $0.7$ eV; there is a clear
reduction of the critical scale (and thus the transition temperature) when
there is a strong competition between $s$- and $d$-wave channels.

\begin{figure}[t]
\centering
\includegraphics{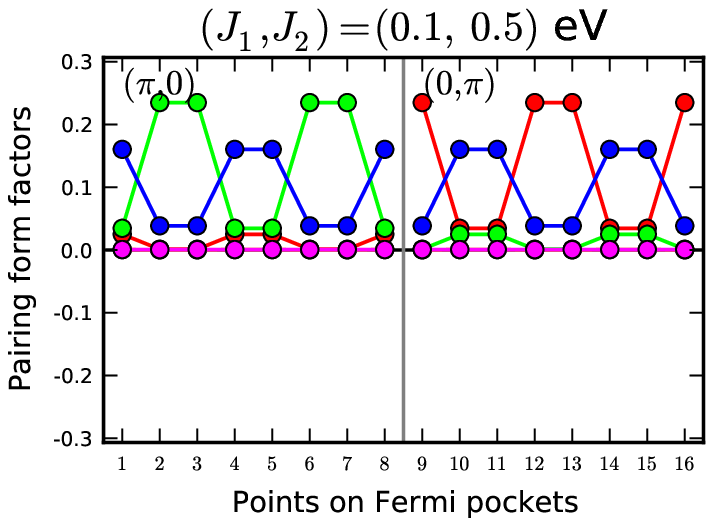}\\%
\vspace{10pt}
\includegraphics{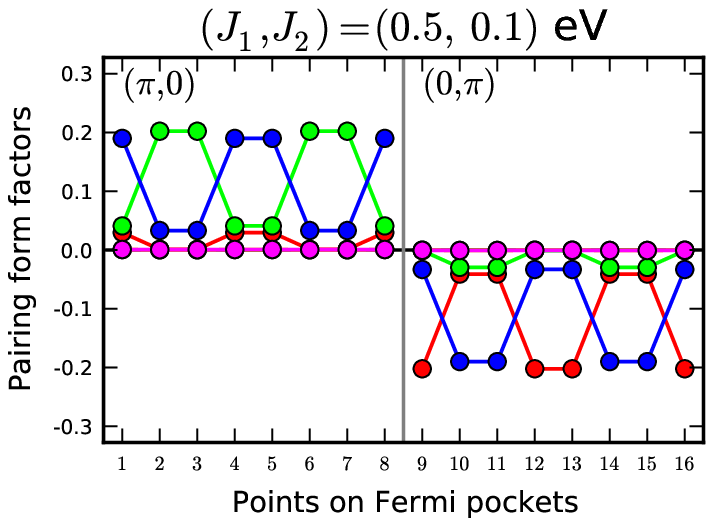}\\%
\caption{\label{fig:orbital-two} (color online) The orbital-resolved pairing form
factors of two typical RG flows. The upper resides in the
dominant $s$-wave and the lower in the $d$-wave regime. The color code is the
same as in Fig.~\ref{bandmaier}, i. e. red $d_{xz}$, green $d_{yz}$,
blue $d_{xy}$, orange $d_{x^2-y^2}$ and magenta $d_{3z^2-r^2}$. }
\end{figure}

Following~\eqref{orbital} we also analyze the orbital decomposition of
the SC pairing from FRG (Fig.~\ref{fig:orbital-two}). We constrain
ourselves to the most relevant three orbitals $d_{xy}$, $d_{xz}$, and
$d_{yz}$.  In particular, we observe that the SC orbital pairing
induces the same sign for all three dominant orbital modes, in correspondence with the mean field
analysis presented before.

\subsection{Three-pocket scenario}
\label{sec:threeps}

Recent ARPES data\cite{sergei} on the chalcogenides may suggest the
existence of a shallow flat pocket around the $\Gamma$ point (the location,
and especially the $k_z$ position of such a pocket are still under debate). By tuning parameters, we have obtained in the previous section a three-orbital model that has an additional electron-pocket around $M$-point in the unfolded BZ.
%The third electron pocket that appears in the
%three-band model employed for the mean-field analysis could be
%seen as an artefact stemming from the 3-orbital approximation.

\begin{figure}[t]
\centering
\includegraphics{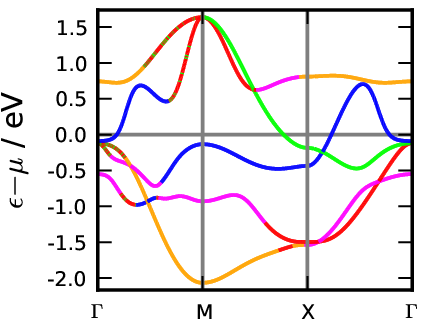}%
\includegraphics{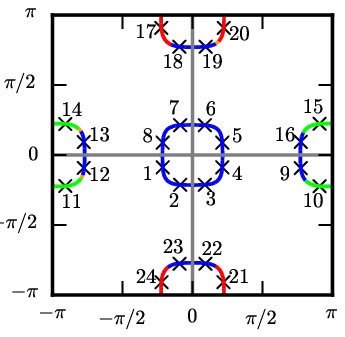}%
\caption{\label{fig:band-three} (color online) The band structure and the Fermi
surfaces of the modified band structure, colored according to the dominant
orbital content. The color code is red $d_{xz}$, green $d_{yz}$, blue $d_{xy}$, orange $d_{x^2-y^2}$ and
magenta $d_{3z^2-r^2}$. The numbered crosses show the center of Fermi surface
patches used in the FRG calculations.}
\end{figure}

In our FRG approach we can take a more profound microscopic
perspective on this issue. From
the true band structure calculations at hand for the chalcogenides,
we consider it unlikely that it will be a hole band regularized up
towards the
Fermi surface.
\begin{figure}[t]
\includegraphics[width=8cm]{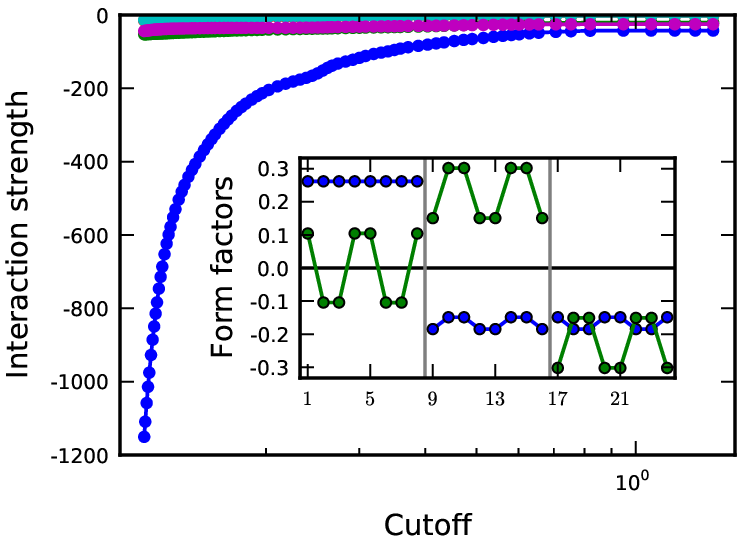}
\caption{\label{fig:j1j2-typical1-three} (color online) Typical RG flows and the
superconducting gaps associated with Fermi surface for the three-pocket
scenario with $(J_1,J_2)=(0.2,0.8)$~eV. Leading form
factor is denoted in blue, sub-leading form factor in green. }
\end{figure}
\begin{figure}[t]
\includegraphics[width=8cm]{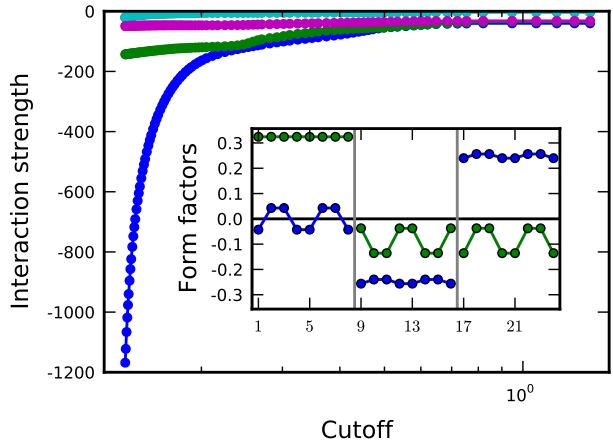}
\caption{\label{fig:j1j2-typical2-three} (color online) Typical RG flows and the
superconducting gaps associated with Fermi surface for the three-pocket
scenario with $(J_1,J_2)=(0.9,0.3)$~eV. Leading form
factor is denoted in blue, sub-leading form factor in green. }
\end{figure}
Instead, we investigate the effect of a possible
electron band at the $\Gamma$ point in the unfolded Brillouin
zone. This is suggested from the folded 10-band calculations, where
one electron-type band closely approaches the Fermi level around
the $\Gamma$ point\cite{Maier2011}. This band should be very flat
and shallow. From the weak coupling perspective of particle-hole pairs created
around the Fermi surfaces, this will probably have a small effect:
particle-hole pairs will only be created up to energy scales of the
depth of the electron band at the $X$ point below the Fermi surface,
providing some hole-type phase space for the electron band at
$\Gamma$. In a $(J_1, J_2)$ picture, however, this may still
significantly promote scattering along $\Gamma \leftrightarrow X$,
which may further stabilize the $s$-wave phase regime. We have hence
developed a modified band structure designed for this scenario. There,
we have bent down the band dominated by $d_{xy}$ in the
two-pocket model band structure\cite{Maier2011} without changing its band vector
and created an electron pocket around $\Gamma$, accordingly of mainly $d_{xy}$
orbital content (Fig.~\ref{fig:band-three}). The band bending was
achieved by
\[
H\rightarrow H+\sum_\mathbf{k,a,b,s}\xi(\mathbf{k})
c_{\mathbf{k}as}^\dagger u_{a}(\mathbf{k}) u^*_{b}(\mathbf{k}) c_{\mathbf{k}bs},
\]
where $u(\mathbf{k})$ is the eigenvector of the band dominated by $d_{xy}$.
The shift of energy $\xi(\mathbf{k})$ was intentionally
chosen such that the $\Gamma$ pocket exhibits some nesting with the $X$ electron pockets.

\begin{figure}[t]
\centering
\includegraphics{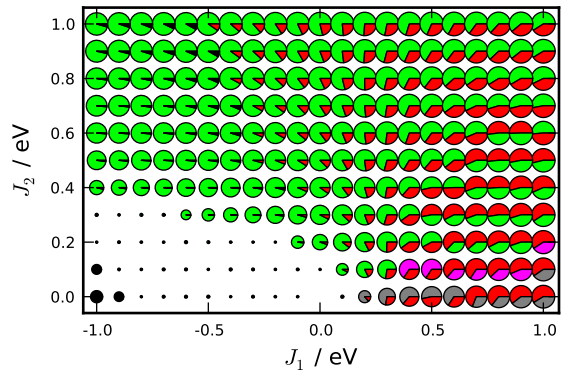}%
\caption{\label{fig:phase-three} (color online) The phase diagram of the three-pocket model.
Here we are able to resolve the $s$-wave
channel into constant $s$-wave (grey), extended $s_{x^2y^2}$-wave (green), and
the nodal $s_{x^2+y^2}$-wave (purple). Parameter sets with $J_1\sim -1$ and
$J_2\sim 0$ have highly oscillating form factors which are due to
artifacts in the calculation; the triplet channel would have to be considered in
these cases.}
\end{figure}

The phase diagram is shown in Fig.~\ref{fig:phase-three}. FRG results for typical
scenario for the $s$-wave and $d$-wave regime are shown in Fig.~\ref{fig:j1j2-typical1-three} and
Fig.~\ref{fig:j1j2-typical2-three}, respectively. As
suspected, the additional pocket strengthens the tendency to form
an $s$-wave in the system, aside from exhibiting an additional
constant $s$-wave instability in a small regime for dominant $J_1$.

\section{Discussion}
\label{sec:dis}

  The above calculations demonstrate that the $s$-wave pairing
symmetry is always robust if the AFM NNN $J_2$ is strong while  a
$d$-wave pairing can be strong if $J_1$ is AFM for the electron
overdoped region. Moreover, if both of them are AFM, there is
strong competition between the $s$-wave
 and $d$-wave pairing. When there are hole pockets, as shown
before\cite{seo2008}, even in a range of $J_1\sim J_2$, the
contribution to pairing from $J_1$ is much weaker than the one from
$J_2$.
 In that case,  an AFM $J_1$ will not generate strong $d$-wave pairing so that the $s$-wave wins easily. From neutron
  scattering experiments, it has been shown that a major difference
  between iron-pnictides and iron-chalcogenides is that the NN
  coupling $J_1$ change from AFM in the former\cite{Zhaojun2008,Zhaoj2009} to FM in the latter\cite{Lip2011}.
  In fact, $J_1$ is rather strongly FM in the latter, which explains
  the high magnetic transition temperature ($500$ K) in
  the 245 vacancy ordering state as shown in Ref.~\onlinecite{Fang2011d}.  Combining these
  results, we can partially answer the
question regarding the different behaviors between iron-pnictides
and iron-chacogenides in the electron-overdoped region: why can the high
SC transition temperature be achieved in the latter, but not in the
former?  Since $J_1$ in iron-pnictides is AFM while it is FM in
iron-chacogenides, $J_1$ will weaken the SC pairing in the former
 but not in the latter.

  A few remarks regarding this work follow: (i) Our mean-field result is qualitatively consistent with the results from a similar model with
  five orbitals\cite{Yu2011,Yu2011mott,yugossi}. The critical difference is regarding
  $J_1$ being FM, and has not been addressed previously; (ii) $s$-wave pairing
  symmetry has also been obtained in Refs.~\onlinecite{
Zhou2011a, Youy2011, wang2011a}. However, the $s$-wave pairing only shows up either
in a narrow region  or with drastically different parameter
settings. Therefore, the $s$-wave is not robust from a microscopic point of
view. Instead, the
$d$-wave is a robust result in these studies. Still, even
the $d$-wave pairing strength based on the scattering between two
electron pockets is generally weak, as specifically
discussed in\cite{Yu2011}, which is another difficulty for this type
of mechanism. (iii) Our results suggest that there is no
difference between iron-pnictides and iron-chalcogenides in terms of
pairing symmetry. Both of them are dominated by $s$-wave pairing. If
both hole and electron pockets are present, the signs of the SC order in
hole and electron pockets are opposite, namely $s^\pm$. However, the
mechanism causing $s^\pm$ is different in the
weak and strong
coupling approach. In the weak coupling approach, the sign change is
due to the scattering between the hole and electron pockets while in
the strong coupling approach, the sign change is due to the form
factor of the SC order parameters which is specified to be $\cos k_x
\cos k_y$ since the pairing mainly originates from the AFM $J_2$. Therefore,
to obtain $s^\pm$ pairing symmetry, the existence of both hole and electron pockets is necessary in the weak coupling approach, but not in the strong coupling one.
(iv) The reason that the superconductivity
vanishes in the iron pnictides in the electron overdoped region is
not solely due to the competition between $s$-wave and $d$-wave pairing
symmetry. It is also due to the weakening of local magnetic exchange
coupling themselves and the reduction of  band width renormalization.
(v) The prospective experimental confirmation of $s$-wave pairing symmetry in $\text{KFe}_2\text{Se}_2$ will
support that superconductivity in iron-based
superconductors might be explained by local AFM exchange couplings. (vi)
Neutron scattering also suggests that there is significant AFM
exchange coupling between two third nearest neighbor sites, i.e.
$J_3$\cite{Lip2011,Fang2009b}. The existence of $J_3$ will further
enhance the $s$-wave pairing since it generates pairing form factors
as $ \cos2k_x +\cos 2k_y$ in reciprocal space which in turn can enhance the
pairing at the electron pockets.

\section{Conclusion}
\label{sec:con}

In summary,  we have shown that the pairing symmetry in
electron-overdoped iron-chalcogenides
is a robust $s$-wave. The fact that the NN magnetic exchange coupling
is FM, which diminishes the possibility of $d$-wave pairing symmetry in these
materials. From a unified perspective of high-$T_c$ cuprates
  and high-$T_c$ chalcogenides, the NN AFM
exchange coupling gives rise to the robust $d$-wave pairing in the  cuprates
while the NNN AFM exchange coupling gives rise to the robust $s$-wave
pairing in iron-based chalcogenide superconductors.

\begin{acknowledgments}
%  JPH would like to thank S. Kivelson, Donglei Feng, Xianhui Chen,
%  Hong Ding, Xiang Tao and Qimiao Si for useful discussions and thanks
%  the Institute of Physics, CAS for research support. We thank
% , S. Graser, P. Hirschfeld, C.~Platt, W. Hanke,
%  D. Scalapino, and J.  van den Brink for discussions.
%
  We thank S. Borisenko, J. van den Brink, Xianhui Chen, A. Chubukov, Hong Ding, Donglei
  Feng, S. Graser, W. Hanke, P. Hirschfeld, S. Kivelson, C. Platt, D.
  Scalapino, Qimiao Si, Xiang Tao, Fa Wang, and Haihu Wen
   for useful discussions. JPH thanks the Institute of Physics, CAS for
  research support.  RT is supported by DFG SPP 1458/1 and a Feodor
  Lynen Fellowship of the Humboldt Foundation and NSF DMR-095242. BAB
  was supported by Princeton Startup Funds, Sloan Foundation, NSF
  DMR-095242, and NSF China 11050110420, and MRSEC grant at Princeton
  University, NSF DMR-0819860.
\end{acknowledgments}

\end{document}